# Early warnings are too late when parameters change rapidly

Rohit Radhakrishnan[1,2], Induja Pavithran[1,2,a)], Valerie Livina[3], Jürgen Kurths[4,5] and R. I. Sujith[1,2*]

[1] Department of Aerospace Engineering, Indian Institute of Technology Madras, Chennai, 600036, India
[2] Centre of Excellence for studying Critical Transitions in Complex Systems, Indian Institute of Technology Madras, Chennai 600036, India
[3] Data Science Department, National Physical Laboratory, Hampton Road, Teddington, TW11 0LW, UK
[4] Potsdam Institute for Climate Impact Research (PIK), Potsdam 14473, Germany
[5] Institute of Physics, Humboldt Universität zu Berlin, Berlin 12489, Germany
[a)] Present address: The Jacob Blaustein Institutes for Desert Research, Ben Gurion University of the Negev

[*]Corresponding author Email: sujith@iitm.ac.in and juergen.kurths@pik-potsdam.de

## Abstract

Early warning signals (EWSs) forewarn a sudden transition (or tipping) from a desirable state to an undesirable state. However, we observe that EWSs detect an impending tipping past bifurcation points when control parameters are varied fast; this questions the applicability of EWSs in real-world systems. When a control parameter is changed at a finite rate, the tipping is also delayed, providing a borrowed stability (in the parameter space) before the system tips. In this study, we use the Hurst exponent as EWS in a thermoacoustic system - a horizontal Rijke tube. We find that upon receiving an EWS alert, a quick reversal of the control parameter within the region of borrowed stability cannot always prevent tipping in real-world systems. We show this failure is due to the (i) delay in receiving the EWS alert and (ii) dispersion observed in the warning points received. For fast variation of parameters, where preventive measures fall short, we demonstrate EWS-based control actions to rescue the system after tipping. Our results in a real-world system for a fast variation of parameter highlight the limits of applicability of EWSs in preventing tipping.

**Introduction**

Tipping point refers to a critical moment or a threshold of control parameter at which a system undergoes a sudden and often irreversible change in its behavior (1-3). Such a sudden transition from a desirable state to an undesirable one can result in catastrophic changes, affecting natural environments, human societies and economies. Several generic early warning signals (EWSs) are developed to forewarn such an impending tipping (4-8).

The primary focus of the research on EWSs involves the identification of indicators across various systems to predict tipping (9). In the case of bifurcation induced tipping, the responses to perturbations slow down on approaching a bifurcation point, a phenomenon known as critical slowing down (CSD). EWSs developed to capture CSD, such as autocorrelation (4) and variance (5) of the system variable are shown to increase prior to shifts in ecosystems (*9*), climate system (10), financial systems (11), lake food webs (12), and other dynamical systems. However, obtaining reliable EWSs in real-world systems is often challenging. Inherent fluctuations or noise can lead to false alarms (13,14). To use EWS wisely, in practical systems, we need to quantify the reliability (15) and sensitivity (16) of EWS for each system. Focusing on developing more general resilience-based EWSs could be one possible solution (17).

Even if we get a reliable EWS alert, implementing appropriate interventions needs sufficient warning time before tipping. Pavithran and Sujith (2021) found that the rate at which the parameters vary affects both the available warning time and the tipping point (18). Parameters in natural and engineering systems evolve continuously. A finite rate of change of the bifurcation parameter results in rate-dependent tipping-delay (18-20); i.e., tipping occurs after crossing the bifurcation point ($\mu_0$) in the case of bifurcation induced tipping (Fig. 1a, b). We notice that early warning for tipping is also delayed from the bifurcation point for fast variation of parameters. Receiving a warning after $\mu_0$ indicates that the system would eventually tip, even if we stop varying the parameter after receiving the warning. A potential strategy to prevent the system from tipping is to promptly revert the control parameter quickly to a lower value (Fig. 1c), within the tipping-delay (19,21). Implementation of such an intervention needs to be thoroughly investigated before applying to real systems. For instance, using a model of epidemic spread, Ullon and Forgoston (2023) showed that control actions performed at the wrong time can even lead to further extension in the time spent in the undesirable region, than if no control had been applied (22). Thus, we need to perform the right action at the right time to prevent the system from tipping, for a given rate of change of parameter.

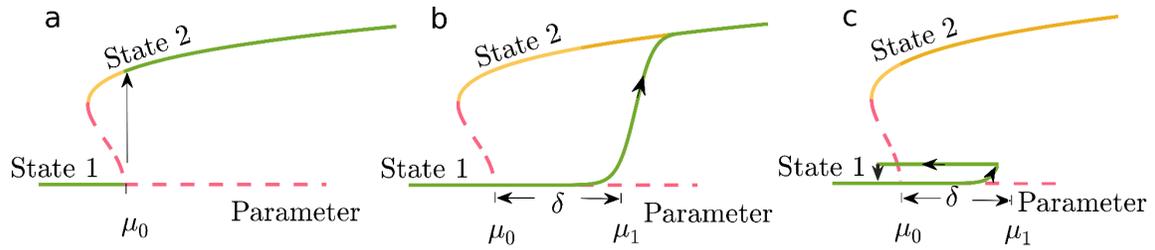

**Fig. 1.** Visualization depicting the transition under different conditions: **(a)** In quasi-static conditions, the system transitions from state 1 to 2 at the bifurcation point $\mu_0$. **(b)** Control parameter changes at a finite rate ($r$). As the system changes from state 1 to 2, during which the control parameter reaches $\mu_1$, after a delay $\delta$ from $\mu_0$. **(c)** Within the parameter range $\delta$, reversing the control parameter can prevent the system from tipping to state 2. (Green curves are representative trajectories in these three different scenarios)

In an experimental system, we design and demonstrate real-time prevention strategies based on the available warning time and the rate at which the control parameter varies. In the current study, we conduct experiments on a prototypical thermoacoustic system, a horizontal Rijke tube (Fig. 2a and see Methods for details of the experiment). As we vary the control parameter, a positive feedback is established between the sound waves in the duct and the unsteady heat release rate from the heated mesh. The positive feedback results in a transition from a state of quiescent operation (low amplitude aperiodic fluctuations) to a highly undesirable state of thermoacoustic instability (TAI), characterized by self-sustained high-amplitude periodic pressure oscillations (Fig. 2b, c). In the context of rockets and gas turbine engines, TAI can lead to structural damage of components (23,24), failure of electronics in the vehicle and the satellite payload (24), excessive heat transfer leading to failure of thermal protection system (24,25), disruptions to guidance and navigation systems, and even result in mission failures (24,26).

Several EWSs (18,27) are reported to give warning before the system goes to TAI, among which any EWS with low false warning rate or combination of EWSs can be chosen for initiating prevention action. Autocorrelation (27,28) and Hurst exponent ($H$) (18,29) are good EWSs for the current system (18). For the present study, we employ $H$ as the EWS (see Methods). Variation in the value of $H$ captures the fractal characteristics of time series signals acquired from the system (18,30) (Fig. 2d).

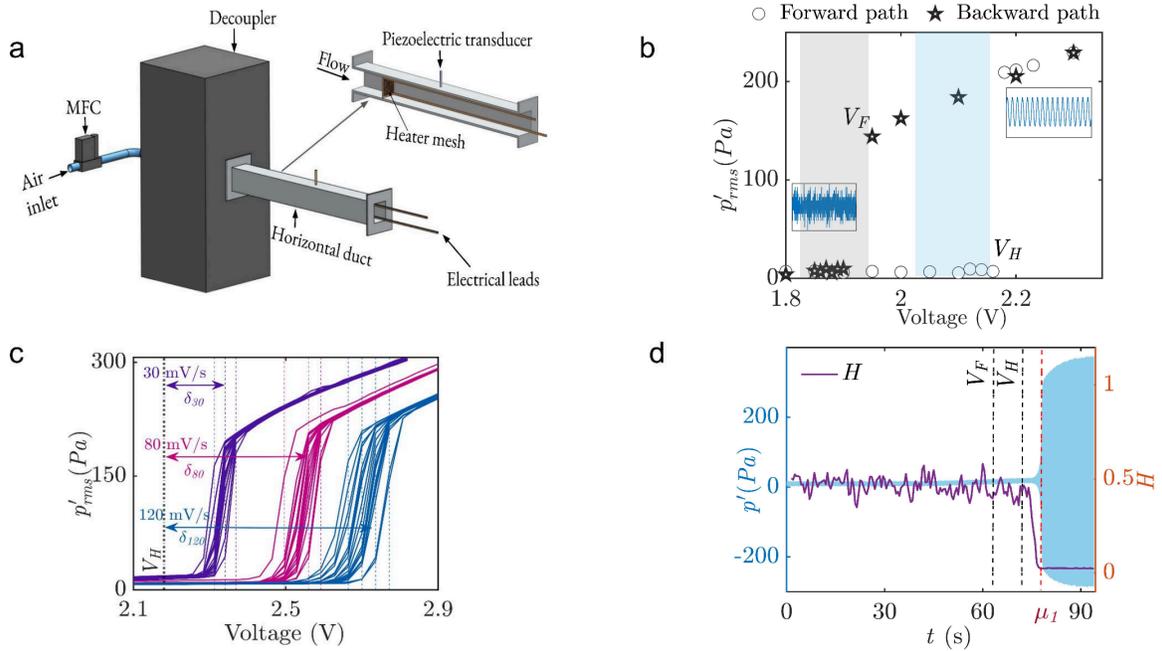

**Fig. 2.** System behavior in response to changes in control parameter (voltage supplied to the heater). **(a)** Schematic of the horizontal Rijke tube comprising a duct, an electrically heated wire mesh, and a decoupler, with a natural frequency of 162 Hz. **(b)** Variation of root mean square of acoustic pressure fluctuations ($p'_{rms}$) for a quasi-static experiment. ○ represents the forward path, with an abrupt jump of $p'_{rms}$ at the Hopf point $V_H$ (2.18 V), while ☆ represents the backward path, with a sudden drop to quiescent state at the fold point $V_F$ (1.95 V). Such a sudden shift from a quiescent state to TAI (i.e., fixed point to limit cycle shown in inset) is a manifestation of Andronov Hopf bifurcation (*31-33*). Shaded regions represent $V_H$ and $V_F$ variations for 15 experimental realizations. **(c)** Tipping points are estimated by finding the maxima of $dp'_{rms}/dt$. The standard deviation in the tipping points is quantified using 20 realizations of experiments for 3 different rates (*r*). The onset of tipping ($\mu_1$) is delayed by $\delta$ from $V_H$. As shown for rates 30 mV/s, 80 mV/s, and 120 mV/s, $\delta$ increases with increase in rate (*18*). **(d)** The variation in the value of *H,* along with the acoustic pressure fluctuations time series as we vary the parameter linearly in time. The value of *H* decreases providing an early warning of an impending tipping well before $\mu_1$.

## Results

We classify the values of rate of change of parameter (*r*) for which the EWS alerts are received before and after crossing the Hopf point as slow and fast variations of change of parameter, respectively. In Fig. 3(b), we vary the control parameter at *r* = 2 mV/s. Upon receiving the EWS alert, we stop the increase of control parameter and freeze its value; we term this prevention strategy as "freeze strategy". Figure. 3c shows the variation in the value of *H* for 20 realizations of experiment at *r* = 2 mV/s. As the EWS alerts are received ahead of the Hopf point, we consider *r* = 2 mV/s as a slow variation of parameter. We observe variability in the timing of EWS alerts (ranging from 906 to 1007 s) across multiple realizations of the experiment. Unni et al. (*34*) reported that inherent fluctuations in a physical system can induce variability in the transition to an oscillatory state during individual experimental realizations.

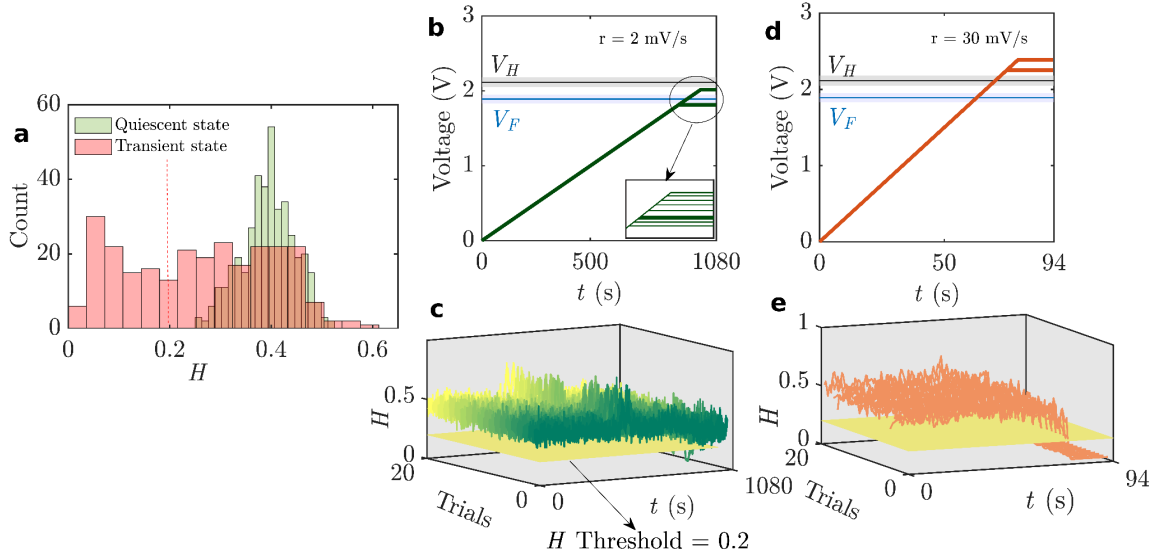

**Fig. 3.** Experimental demonstration of freeze strategy for a slow variation of parameter and failure of this prevention strategy under fast variation of parameter. **(a)** Distribution of values of $H$ from multiple trials for the quiescent operation state (at 1.84 V; green) and the transient state (from 1.84 V till the onset of tipping; red) indicates that a value of 0.2 for $H$ can distinguish these two states. **(b)** The voltage variation for 20 trials at $r$ = 2 mV/s is shown in the inset, out of which the cases where we received the maximum and minimum values for EWS alerts shown in thick solid lines. **(c)** Upon receiving an EWS alert, the "freeze strategy" proves to be an effective prevention strategy, as we observe that the value of $H$ returns back to higher values after the prevention strategy (green shades depict different trials). **(d)** For $r$ = 30 mV/s, we observe that the maximum and minimum values for EWS alerts are received above $V_H$. **(e)** For the case of fast variation of parameter $r$ = 30 mV/s, the freeze strategy is not a viable solution, as observed from the value of $H$ approaching zero (orange shades show different trials) indicating the presence of periodic content in the signal even after prevention strategy.

Values of $H$ during the quiescent state for the Rijke tube lie within the range of 0.4 to 0.6. After the implementation of prevention strategy, the values of $H$ return back to 0.4 to 0.6 (Fig. 3c). Thus, the "freeze strategy" proves to be an effective prevention strategy for a slow variation of parameter. However, for a fast variation of parameter $r$ = 30 mV/s (Fig. 3d), the freeze strategy could not prevent the transition to TAI, observed from the value of $H$ approaching zero (Fig. 3e) indicating the presence of periodic content in the signal (i.e., the system transitioned to TAI). Therefore, we need to adopt an alternative prevention strategy for fast variation of parameters.

For $r$ = 30 mV/s, the EWS alert is received after the control parameter crosses the Hopf point. Therefore, upon receiving an EWS alert, we abruptly reduce the control parameter to a value of 1.96 V, which is lower than $V_H$ but still within the bistable region; therefore, this action qualifies as a prevention strategy (see Methods). Out of 100 trials, we observe a success rate of 76% in preventing tipping (Fig. 4a, b) with the strategy of direct cut-off of voltage to a value of 1.96 V within the bistable region. Failures in the remaining 24 cases are attributed to delayed EWS alerts received (between 2.39 V to 2.41 V) compared to successful cases (between 2.31 V to 2.37 V), resulting in insufficient time for prevention action. Subsequently, for $r$ = 80 mV/s we perform the prevention strategy of direct cut-off of voltage to a value within the bistable region (Fig. 4c, d). Out of 100 trials, prevention strategy was successful only in six trials. In the remaining 94 trials, the EWS alert is received at voltage values (between 2.54 V to 2.66 V) that are delayed compared to successful cases (between 2.46 V to 2.5 V), providing insufficient time for intervention and the system transitions to periodic oscillations. In spite of the borrowed stability provided by the delay in tipping due to rate, with increase in $r$, the EWS alerts are also received after crossing the Hopf point. As a consequence, even extreme prevention strategies, such as directly reducing the control parameter to a lower value within the bistable region, may be rendered ineffective.

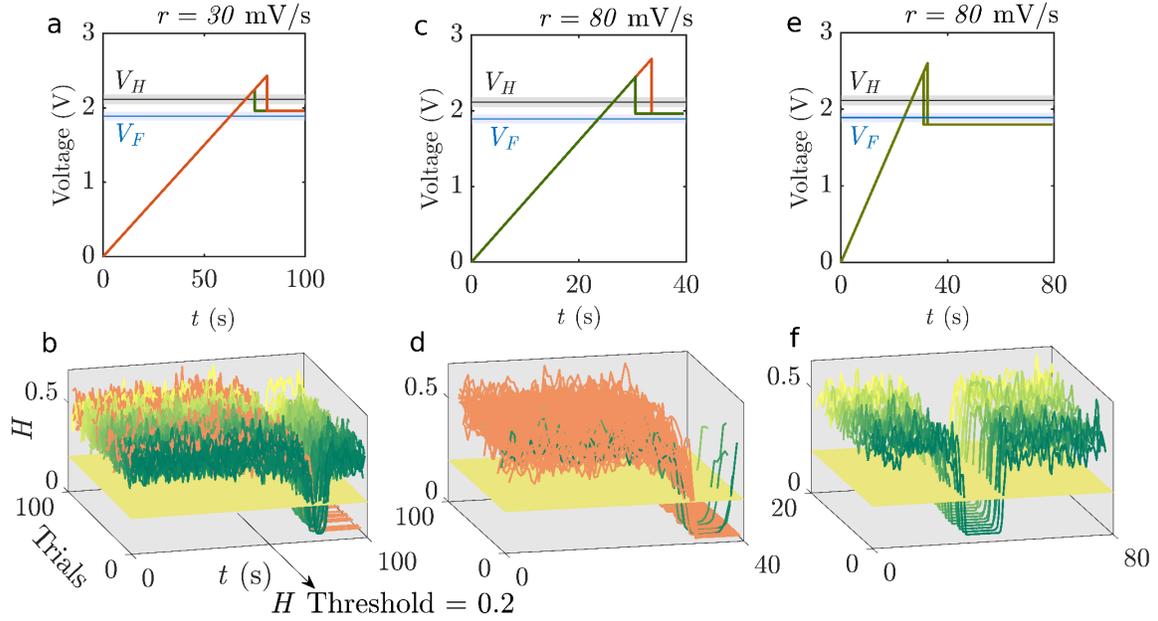

**Fig. 4.** EWS-based prevention strategy of direct cut-off to a lower value of control parameter within the bistable region for fast variation of parameter and alternative solution of control action: **(a,c,e)** The rate of variation of voltage illustrated for cases with lowest and highest EWS alert value for $r$ = 30 mV/s, 80 mV/s and 120 mV/s is depicted. **(b)** For $r$ = 30 mV/s, out of 100 realizations performed, the strategy of direct cut-off of voltage to a value of 1.96 V which lies within the bistable region successfully prevents tipping in 76% of trials (gradients of green color indicates these trials), while in the remaining 24% (gradients of orange color indicates these trials) EWS alerts are received farther away from Hopf point ($V_H$), resulting in tipping to TAI even after prevention action. **(d)** For $r$ = 80 mV/s, we observe that EWS alerts are received at higher values of voltage compared to $r$ = 30 mV/s case. We could prevent the transition only in 6 realizations, with the prevention strategy of direct cut-off of voltage to 1.96 V. **(f)** In the case of $r$ = 80 mV/s, where prevention strategy proves ineffective, lowering the control parameter to a value below the fold point allows us to bring back the system to its quiescent operational state. The system exhibits periodic oscillations for a finite duration before reverting to its quiescent operational state.

When the prevention strategy (Fig. 4c, d) falls short, we need to adopt an alternate approach. Abruptly lowering the control parameter below the fold point of the system (Fig. 4e, f) effectively controls the system after tipping; we refer to this action as control action (see Methods). The control action returns the system to the quiescent operating state (increase in $H$ from 0 to 0.4 - 0.6 in Fig. 4f). For fast variation of parameter, where extreme prevention strategy fails, the control action emerges as the sole viable strategy for restoring the system to the state of quiescent operation. The control action can be useful in real-world systems that can tolerate oscillations for a short period.

**Discussion**

We have investigated the feasibility of EWS-based prevention strategies in a real-world system - an experimental thermoacoustic system. We demonstrated that the delay in the onset of tipping increases with increasing the rate of change of parameter (Fig. 2c). We showed that for a slow variation of parameter, the freeze strategy upon receiving the EWS alert is an effective prevention strategy. For fast variation of parameter, directly lowering the control parameter to a value within the bistable region proves to be an effective prevention strategy. Practical systems often exhibit variabilities in stability boundaries, onset of tipping points, and EWS-alert timings. Therefore, considering these variabilities is crucial while performing real-time prevention actions in real-world systems. For fast variation of control parameter, EWS-alerts are received after crossing the Hopf point, rendering prevention strategy less effective due to limited time for intervention. We have demonstrated that in cases where prevention strategy fails, control actions can be employed in systems where periodic oscillations are tolerable temporarily. We leave the choice of EWS to the practitioners since different EWSs work well for different systems. The current study serves as the first experimental study exploring the reliability and limits of applicability of EWS in preventing tipping in real-time for a practical system.

**Materials and Methods**

**Quasi-static experiments**

We perform experiments in a prototypical thermoacoustic system known as a horizontal Rijke tube (Fig. 2a). A horizontal Rijke tube is a 1m long duct of cross-sectional area 0.093 x 0.093 $m^2$. An airflow of 100 SLPM (2.04 g/s) is maintained throughout the experiments. The duct has two ends: one end is open to the atmosphere, and the other is connected to a decoupler, which is a large rectangular chamber. The decoupler (1.2 m × 0.45 m × 0.45 m) removes the inherent fluctuations in the incoming air before the air enters the Rijke tube. A wire mesh located 25 cm from the decoupler end is electrically heated, which acts as a compact source of heat in the Rijke tube. A piezoelectric pressure transducer (PCB103B02, sensitivity: 247.5 mV/kPa, resolution: 0.2 Pa, and uncertainty: 0.15 Pa) is mounted on the duct at a location 57 cm (i.e. near the pressure antinode of the fundamental duct mode) from the decoupler end to measure the acoustic pressure fluctuations in the system. We acquired the acoustic pressure fluctuations signal for each value of the control parameter for 6 s at a sampling frequency of 10 kHz (see Fig.2b). A bin size of 0.17 Hz is used for obtaining the fast Fourier transform (FFT) of the acoustic pressure fluctuations time series. The fundamental frequency of the system is nearly 162 Hz. Prior to the experiments, under cold flow conditions, a sinusoidal signal is input to the Rijke tube using a loudspeaker, and then the loudspeaker is switched off abruptly. The acoustic pressure fluctuation decays when the loudspeaker is switched off, and by performing the Hilbert transform and calculating the logarithmic decay of acoustic pressure fluctuations data acquired, we determine the decay rate. A decay rate of $13 \pm 0.5$ s$^{-1}$ is maintained for all the experiments conducted, to ensure that the experiments can be repeated.

When we increase the control parameter (referred to as the forward path in Fig.2b), we observe the root mean square (RMS) value of the acoustic pressure fluctuations ($p'_{rms}$) is close to zero until a certain range of voltage values. As we increase the voltage, the system changes from a dynamical state known as fixed point (quiescent operation) to a state of limit cycle oscillations (high amplitude periodic oscillations) at a particular value of the control parameter. We observe a sudden rise in $p'_{rms}$, a characteristic of subcritical Hopf bifurcation (18,27) and the voltage value corresponding to this sudden transition is considered as the value at the Hopf point ($V_H$). Subsequently, as we decrease the voltage (reverse direction), limit cycle oscillations display a steady decrease in amplitude. However, when we reverse the direction of control parameter, the transition from limit cycle to a fixed point occurs at the voltage value corresponding to the fold point ($V_F$), which is less than the $V_H$. We observe a hysteresis region, which is a characteristic of subcritical Hopf bifurcation.

**Selection of threshold for the Hurst exponent**

In the current study with the acquired acoustic pressure fluctuations signal, we compute the Hurst exponent ($H$) using the multifractal detrended fluctuation analysis (35) (MFDFA). In MFDFA, the value of $H$ is calculated from the scaling of the root mean square (RMS) of the standard deviation of acoustic pressure fluctuations in relation to the data segment length. For the experiments with a finite rate of change of parameter, we compute the Hurst exponent with a window of 0.4 s data, overlapped by 0.2 s data. Our choice of window size is deliberate, as it strikes a balance between the efficient computation of $H$ and the ability to issue timely warnings with only 0.2 s of data, affording more time for prevention action.

To determine the threshold for $H$, we acquired the acoustic pressure fluctuations for a quiescent operation state (below the fold point at 1.84 V) and corresponding values of $H$ are depicted in (Fig. 3a) as a histogram. To ascertain the transition state, we conducted multiple realizations, allowing the system to evolve from a quiescent state to a state of periodic oscillation. We then selected the values of $H$ for the pressure fluctuations within the control parameter range from 1.84 V to the onset of tipping.

We aim to identify an optimal threshold that minimizes false alarms while providing sufficient time for prevention actions. Depending on the risk tolerance and specific requirements of the operator, the threshold can be tailored to initiate the necessary prevention strategy.

When the value of $H$ crosses the threshold selected, we provide an EWS alert to initiate the prevention action. For a rate of change of parameter of 30 mV/s, the EWS alerts received for 20 trials each with different thresholds are depicted in Fig. 5. We observe that for a threshold of 0.3, the EWS alerts for 20 trials are received at very low voltage values compared to the voltage value at the onset of tipping ($u_1$), giving false warnings (as shown in Fig. 5, blue circles). When we set the threshold to 0.2, as well as the threshold to 0.1, we find that the voltage corresponding to the EWS alert is close to $\mu_1$. However, the earlier the warning is received, the earlier we can perform the required prevention actions, to prevent the system from tipping. Hence, an EWS threshold of 0.2 is selected for initiating the prevention actions.

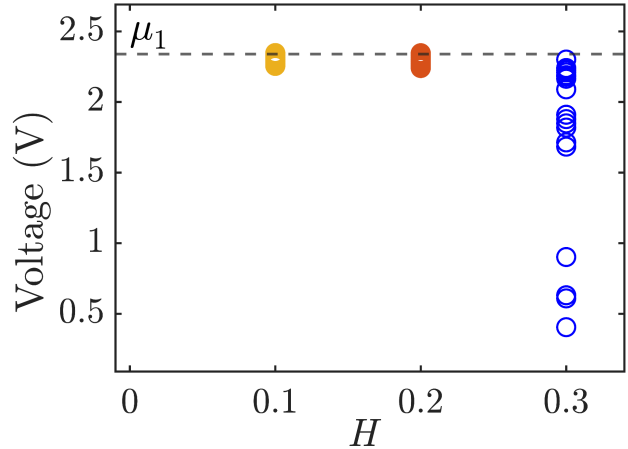

**Fig. 5.** Variation in the EWS alerts received for different thresholds: For a rate of 30 mV/s, the voltage at which the Hurst exponent crosses the threshold is plotted against the corresponding threshold for 20 realizations of experiment each with different thresholds. For a threshold of 0.3, alerts are received before the tipping point ($\mu_1$). However, the variation across different trials is very high and includes many false warnings. For thresholds of 0.2 and 0.1, the alerts are received closer to $\mu_1$. A threshold value of 0.2 for $H$ is found to be an optimum value with minimum false alarms and maximum warning time.

**Receiver operating characteristics (ROC)**

For the present study, ROC curves quantify the ability of $H$ in distinguishing the quiescent operation state (considered positive) and the transition to thermoacoustic instability (negative). We determine the,
True positive (TP) = cases rightly classified as positive
True negative (TN) = cases rightly classified as negative
False positive (FP) = cases actually negative but classified as positive
False negative (FN) = cases actually positive but classified as negative.
Then we calculate the false positive rate (FPR) = FP / (FP+TN) and the true positive rate (TPR) = TP / (TP+FN). The FPR is plotted against the TPR for different threshold values for 3 different rates of change of parameter (shown in Fig.6). For all the 3 cases, the curves remaining close to the top-left corner indicate that the classifier performs well (15,36). We also compute the area under the ROC curve (AUC) and observe values close to one indicating perfect distinction performed by the classifier.

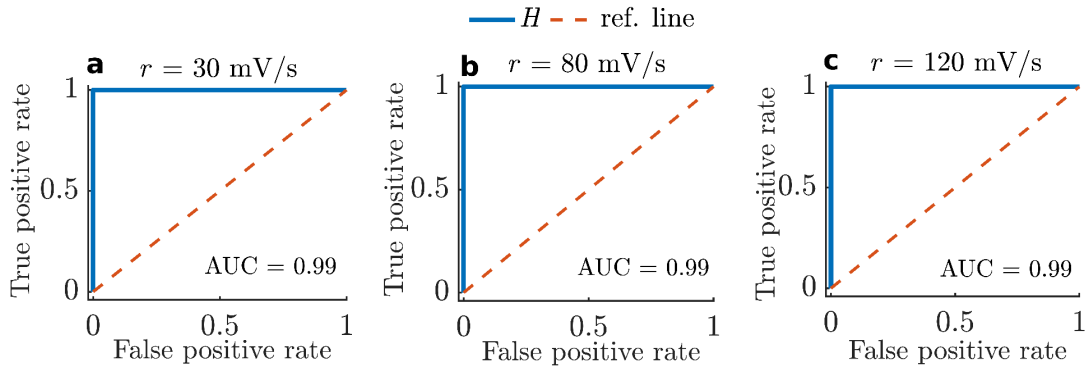

**Fig. 6.** ROC curves plotted using the values of *H* observed for rate of change of parameter (a) 30 mV/s (b) 80 mV/s (c) 120 mV/s. The area under the curve is nearly 0.99 for all cases indicating a perfect distinction performed by the classifier, compared to a random classification (depicted with dotted lines).

**Intervention strategies - prevention and control actions**

We perform the following intervention strategies upon receiving an EWS alert:

1) Prevention strategy - For fast variation of control parameters, tipping occurs past the bifurcation point. EWS warns of an impending tipping within this tipping delay. However, the system will exhibit tipping if we stop varying the control parameter after receiving the alert, as we already crossed the bifurcation point by the time we get the EWS alert. Therefore, we immediately reverse the direction of variation of the control parameter; we lower the voltage to 1.96 V. We prevent the system from going to TAI by bringing the control parameter below the Hopf point. (Refer Fig. 4 a, b in main text)

2) Control strategy - When an EWS alert is obtained near the tipping point under fast variations, we have less warning time than slow parameter variations. In such scenarios, the system undergoes tipping while we reverse the control parameter. Here, reducing the voltage to a value lower than the Hopf point is not enough to prevent the system from tipping to TAI. Therefore, we reduce the control parameter below the fold point so that the system returns to the quiescent state immediately after the tipping. Thus, we call this a control strategy for mitigating TAI. We utilize such a technique in scenarios where the prevention strategy fails. (Refer Fig. 4 e, f in main text)

**Quantification of the abrupt prevention actions**

We define the window size for calculating the *H*, such that the value of *H* converges to a fixed value at different states of the system. We acquired the time series of acoustic pressure fluctuations at each state of the system under quasi-steady experiments. The data was acquired at a sampling frequency of 10 kHz for a period of 6 s. Figure 7 indicates that values of *H* converge to a fixed value for window size of 50 cycles and above. We selected 64 acoustic cycles as the window size for calculating *H*. For the experiments with a finite rate of change of parameter, we use a moving window of 0.4 s data with an overlap of 0.2 s data. The total time for calculating the *H*, comparing the value of *H* with the threshold and sending the signal to the DC power supply takes time of the order $10^{-4}$ s (clock time in Matlab).

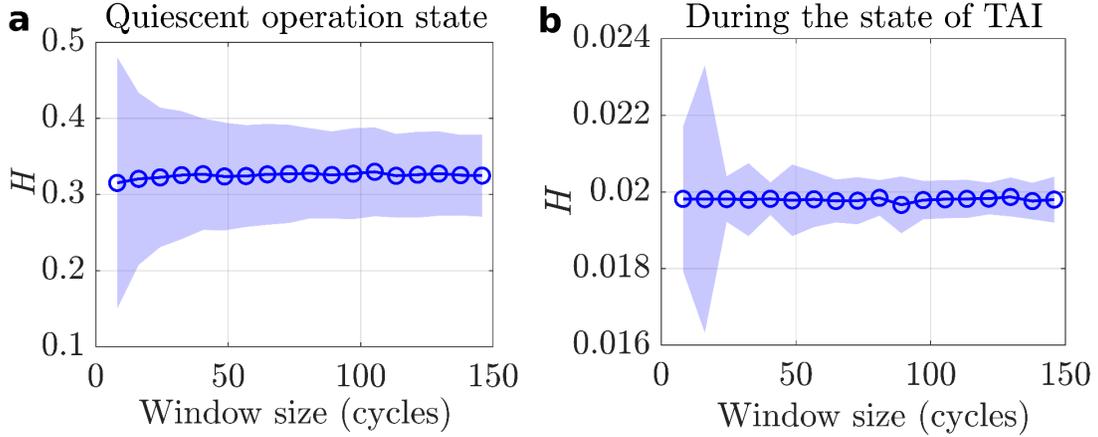

**Fig. 7.** Variations in the value of *H* for different moving window sizes at different states of operation in Rijke tube: The Hurst exponent with standard deviation corresponding to the quasi-steady time series of acoustic pressure fluctuations acquired at a sampling frequency of 10 kHz for a duration of 6 s with the size of the moving window for the states of (a) quiescent operation (b) during the state of TAI is plotted for 14 quasi-static experiments. For the window size of 50 acoustic cycles and above the value of *H* tends to a fixed value with small variance. In the current study, we used a window size of 64 cycles.

Once the value of *H* crosses the threshold value of 0.2, we perform the prevention action of direct cut-off to a voltage value of 1.96 V. A programmable DC power supply (TDK - Lambda GEN 8 V - 400 A) is used to control the voltage supplied to the heater mesh. The DC power supply has a response time of 20 ms; i.e., for a given input value of voltage, it takes 20 ms for the DC power supply to reach the aforementioned voltage. Hence, reducing the voltage directly to a voltage value (either to a value within the bistable region or below the fold point) within 20 ms is an abrupt prevention or control action compared to the 0.2 s needed to acquire the next set of data. Since the response time is nearly one order higher than the time required for collecting the data, the prevention action of direct cut-off could be considered as the extreme case of prevention action.

## Acknowledgments

We thank Mr. Ambedkar S. and Mr. Dhadphale J. M for their help during the experiments. RR acknowledges the support from the Prime Minister Research Fellowship, Government of India. We acknowledge the research support in part by the International Centre for Theoretical Sciences (ICTS) for the program "Tipping Points in Complex Systems" (code: ICTS/tipc2022/9). RIS acknowledges the funding from IoE initiative (SP22231222CPETWOCTSHOC). Department of Science and Technology, Government of India. SERB/CRG/2020/003051. VNL was funded by the National Measurement System programme supported by the UK Government's Department for Science, Innovation and Technology.

## Author Contributions

Conceptualization : RR, IP, RIS;Data curation: RR; Formal analysis: RR, IP, RIS; Investigation: RR, IP, RIS, VL, JK; Methodology: RR, IP, RIS, VL, JK; Supervision: IP, RIS, VL, JK; Validation: RR; Visualization: RR, IP; Funding acquisition: RIS; Project administration: RIS; Writing – original draft: RR; Writing – review & editing: IP, RIS, VL, JK

## Competing Interest Statement

Authors declare that they have no competing interests.